\documentclass[runningheads]{llncs}

\usepackage{times}
\usepackage{epsfig}
\usepackage{graphicx}
\usepackage{amssymb}
\usepackage{pifont}
\usepackage{amsmath}
\usepackage{hyperref}
\usepackage{amsmath}
\usepackage{amssymb}
\usepackage{dsfont}
\newcommand{\app}{\raise.17ex\hbox{$\scriptstyle\sim$}}
\usepackage{multirow}
\usepackage{lipsum}
\newcommand\blfootnote[1]{%
  \begingroup
  \renewcommand\thefootnote{}\footnote{#1}%
  \addtocounter{footnote}{-1}%
  \endgroup
}
\makeatletter
\def\blfootnote{\gdef\@thefnmark{}\@footnotetext}
\makeatother

\begin{document}
\title{Characterizing Renal Structures with 3D Block Aggregate Transformers}
\author{Anonymous}
\author{
Xin Yu\inst{1*} \and
Yucheng Tang\inst{2*} \and
Yinchi Zhou\inst{1} \and
Riqiang Gao\inst{1} \and
Qi Yang\inst{1} \and
Ho Hin Lee\inst{1} \and
Thomas Li\inst{3} \and
Shunxing Bao\inst{1} \and
Yuankai Huo\inst{1,2} \and
Zhoubing Xu\inst{4} \and
Thomas A. Lasko\inst{1,5} \and
Richard G. Abramson\inst{6} \and
Bennett A. Landman\inst{1,2,3,6}}
%

\institute{Anonymous}

\institute{Department of Computer Science, Vanderbilt University \and
Department of Electrical and Computer Engineering, Vanderbilt University \and
Department of Biomedical Engineering, Vanderbilt University \and
Siemens Healthineers\and
Department of Biomedical Informatics, Vanderbilt University Medical Center\and
Department of Radiology, Vanderbilt University Medical Center}
\maketitle              
\begin{abstract}
Efficiently quantifying renal structures can provide distinct spatial context and facilitate biomarker discovery for kidney morphology. However, the development and evaluation of transformer model to segment the renal cortex, medulla, and collecting system remains challenging due to data inefficiency. Inspired by the hierarchical structures in vision transformer, we propose a novel method using 3D block aggregation transformer for segmenting kidney components on contrast-enhanced CT scans. We construct the first cohort of renal substructures segmentation dataset with 116 subjects under institutional review board (IRB) approval. Our method yields the state-of-the-art performance (Dice of 0.8467) against the baseline approach of 0.8308 with the data-efficient design. The Pearson R achieves 0.9891 between the proposed method and manual standards, and  indicates the strong correlation and reproducibility for volumetric analysis. We extend the proposed method to the public KiTS dataset, the method leads to improved accuracy compared to transformer-based approaches. We show that the 3D block aggregation transformer can achieve local communication between sequence representations without modifying self-attention, and it can serve as an accurate and efficient quantification tool for characterizing renal structures. \blfootnote{* equal contribution}

\end{abstract}

\begin{figure}[t]
\centering
\includegraphics[width=\textwidth]{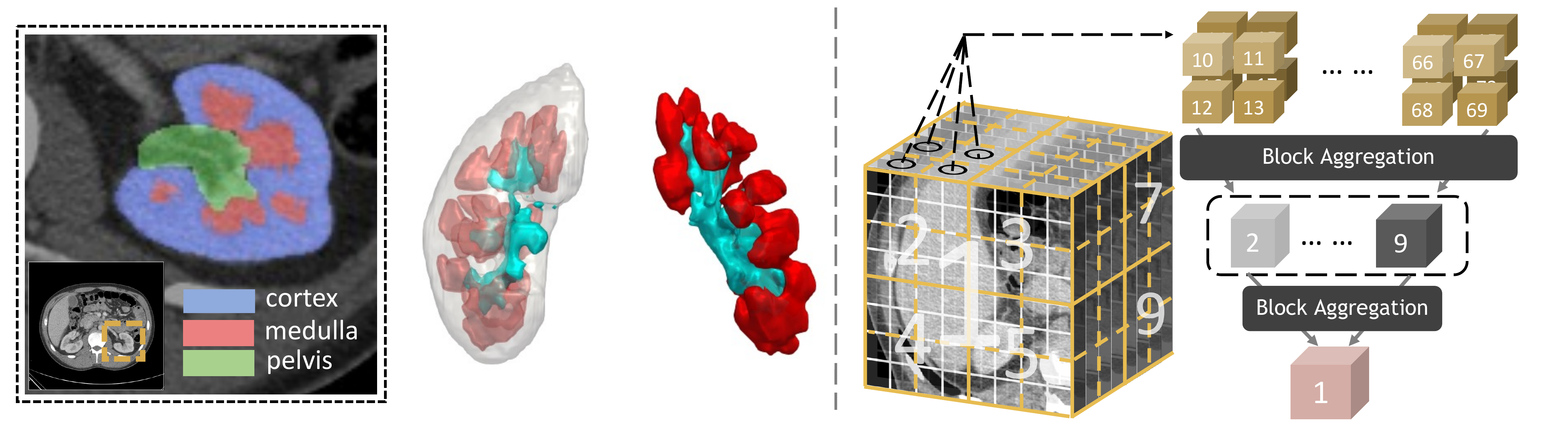}

\caption{Left: visual and 3D illustration of the kidney components. Right: Demonstration of the hierarchical transformer design, the 3D block aggregation is conducted every two hierarchies, blocks at a factor of 8 are merged to perform communication of sequence representations.}

\label{fig:fig1}
\end{figure}

\section{Introduction}
\label{Introduction}

Hierarchical models~\cite{cciccek20163d,roth2018multi,tang2021self} are received significant interest in medical image analysis due to their advantages of modeling heterogeneous high-resolution radiography images. Recent works on vision transformers~\cite{dosovitskiy2020image,liu2021swin} show superior performance on visual representations compared to state-of-the-art convolution-based networks~\cite{he2016deep}. However, ViT usually requires large-scale  training data with expensive clinical expertise~\cite{tang2021self,zhou2021nnformer}. When trained on smaller cohorts, transformer-based models often suffer from a lack of inductive bias~\cite{cordonnier2019relationship,dosovitskiy2020image} and lead to data inefficiency. Moreover, the self-attention mechanism on modeling multi-scale features for high-resolution medical volumes is computationally expensive~\cite{beltagy2020longformer,han2021transformer,liu2021swin}. These challenges inspire designing hierarchical transformer structures in analogy to convolution-based networks (e.g., 3D UNet~\cite{cciccek20163d}). Addressing the data inefficiency of transformers is critical for its application on medical image analysis, especially for understanding small targets on 3D high-dimensional image volumes. 

One such challenge is segmenting the small structures of kidney sub-components. Renal structure volumes from clinical CT scans have been recently suggested as a useful surrogate for evaluating renal function~\cite{lee2003dynamic,van2005functional}. These investigations elucidate the correlations of the volumetric measurements on the renal cortex, medulla, and pelvicalyceal system with kidney function. In such studies, manual segmentation is performed as the gold standard for visual and quantitative morphological assessment on CT scans~\cite{sahani2005multi} as shown in Figure~\ref{fig:fig1}. However, manual quantification by clinical experts is resource-intensive, time-consuming, and may suffer from insufficient inter- and intra-reproducibility.

To improve the representation learning of transformers in small datasets, recent works envision the use of local self-attention to form hierarchical transformers~\cite{liu2021swin,cao2021swin,han2021transformer}. To leverage information across embedded sequences, "shifted window" transformers~\cite{liu2021swin} are proposed for dense predictions and modeling multi-scale features. However, these attempts that aim to complicate the self-attention range often yield high computation complexity and data inefficiency. Inspired by the aggregation function in the nested ViT~\cite{zhang2021nested}, we propose a new design of a 3D U-shape medical segmentation model with Nested Transformers (UNesT) hierarchically with the 3D block aggregation function, that learn locality behaviors for small structures or small dataset. This design retains the original global self-attention mechanism and achieves information communication across patches by stacking transformer encoders hierarchically.  

Our contributions in this work can be summarized as:
\begin{itemize}
\item[\textbullet] We introduce a novel 3D medical segmentation model, named UNesT with a 3D block aggregation function. This method achieves hierarchical modeling of high-resolution medical images and outperforms local self-attention variants with a simplified design, which leads to improved data efficiency.

\item[\textbullet] We collect and manually delineate the first renal substructures dataset (116 patients) on characterizing multiple kidney components. We show that our method achieves state-of-the-art performance to accurately measure the cortical, medullary, and pelvicalyceal system volumes.

\item[\textbullet] We demonstrate the clinical utility of this work by accurate volumetric analysis, strong correlation, and reproducibility. Validation on external public dataset KiTS shows the generalizability of the proposed method.
\end{itemize}

\section{Related Works}
\textbf{3D Medical Segmentation with Transformers.} 
 Transformer-based 3D medical image segmentation models~\cite{wang2021transbts,hatamizadeh2021unetr,xie2021cotr,zhang2021transfuse,jia2021bitr,peiris2021volumetric,zhou2021nnformer,valanarasu2021medical,chang2021transclaw} are popular and achieve state-of-the-art performance in several benchmarks. The self-attention mechanism~\cite{vaswani2017attention} allows the inputs at different positions of a sequence to interact with each other, and then compute the overall representation from the sequences. Although transformers exhibit outstanding performance in learning global context, their deficiency in capturing localized information remain. To address this, TransFuse~\cite{zhang2021transfuse}, TransBTS~\cite{wang2021transbts}, CoTr~\cite{xie2021cotr}, UNETR~\cite{hatamizadeh2021unetr} are proposed architectures which combine transformers and CNNs into hybrid designs. More recently, hierarchical transformers are proposed with shifted-window~\cite{liu2021swin}, it enables cross-patch self-attention connections. Based on Swin ViT,  Swin UNETR~\cite{hatamizadeh2022swin,tang2021self} and SwinUNET~\cite{cao2021swin} are introduced for capturing multi-scale features in CT images. However, the modification on local self-attention results in quadratic increase of complexity.

\noindent\textbf{Hierarchical Feature Aggregation.} The aggregation of multi-level features could improve the segmentation results by merging the features extracted from different layers. Modeling hierarchical features, such as U-Net~\cite{cciccek20163d} and pyramid networks~\cite{roth2018multi}, multi-scale representations are leveraged. The extended feature pyramids compound the spatial and semantic information through two structures, iterative deep layer aggregation which fuses multi-scale information as well as hierarchical deep aggregation which fuses representations across channels. In addition to single network, nested UNets~\cite{zhou2018unet++}, nnUNets~\cite{isensee2021nnu}, coarse-to-fine~\cite{zhu20183d} and Random Patch~\cite{tang2021high} suggest multi-stage pathways that enrich the different semantic levels of feature progressively with cascaded networks. Different from the above CNN-based methods, we explore the use of data-efficient transformers for modeling hierarchical 3D features by the block aggregation.

\section{Method}

\subsection{UNesT Architecture}
\begin{figure}[t]
\centering
\includegraphics[width=\textwidth]{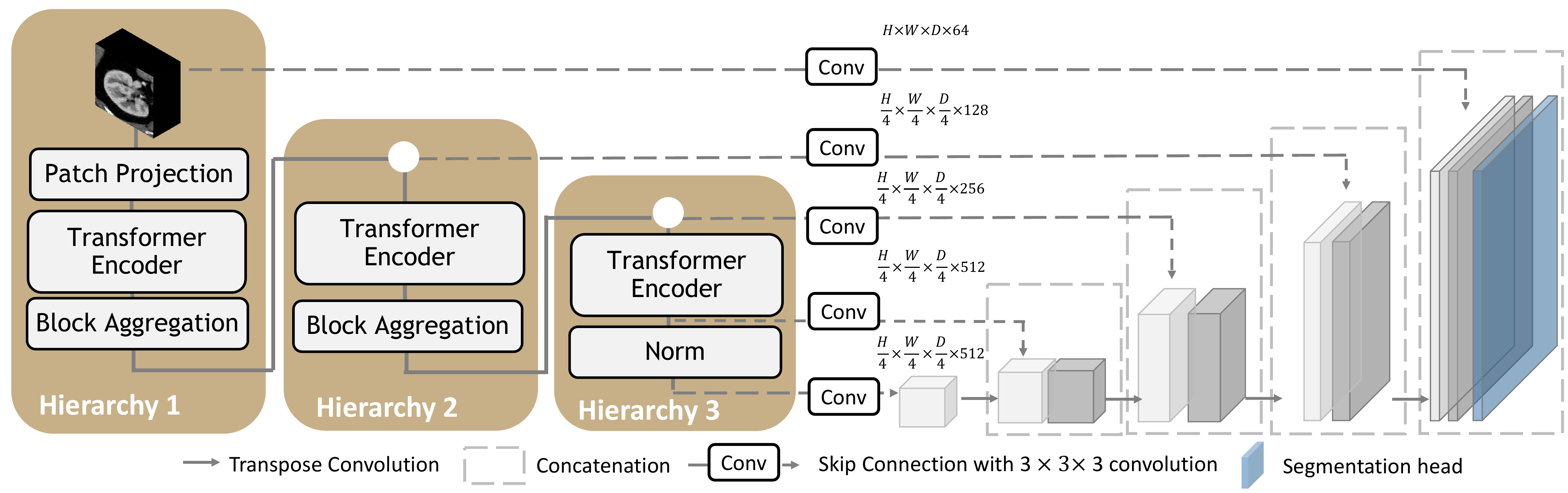}

\caption{Overview of the proposed UNesT with the hierarchical transformer encoder. Block aggregation and image feature down-sampling are performed between hierarchies. }

\label{fig:fig2}
\end{figure}

The proposed network contains a hierarchical transformer as the encoder, which consists of three hierarchies to perform self-attention communications among image blocks. Following the motivation of NesT~\cite{zhang2021nested} for natural images, we process the volumetric information between 3D adjacent blocks by the aggregation layer every two hierarchies. The overall architecture, as shown in Figure~\ref{fig:fig2}, also contains skip connections with convolution modules and a decoder for better capturing localized information. 

Given the input image sub-volume $\mathcal{X} \in \mathbb{R}^{H\times{W}\times{D}}$, the volumetric embedding token is with patch size of $S_h \times S_w \times S_d$. Then all projected sequences of embeddings are partitioned to blocks with a resolution of $\mathcal{X} \in \mathbb{R}^{b \times T \times{n}}$, where $T$ is the number of blocks at the current hierarchy, $b$ is the batch size, $n$ is the total length of sequences. The dimensions of the embeddings follow $T \times{n} = \frac{H}{S_h} \times \frac{W}{S_w} \times \frac{D}{S_d}$. In the subsequent transformer layers, we use the canonical multi-head self-attention (MSA), multi-layer perceptron (MLP), and Layer normalization (LN). We add learnable position embeddings to sequences for capturing spatial relations before the blocked transformers. The output of encoder layers $t-1$ and $t$ are computed as follows:

\begin{equation}
\begin{array}{l}
\hat{{z}}^{t}=\text{MSA\textsubscript{HRCHY\textsubscript{l}}}(\text{LN}({z}^{t-1}))+{z}^{t-1} \\
{z}^{t}=\text{MLP}(\text{LN}(\hat{{z}}^{t}))+\hat{{z}}^{t}, \\

\end{array}
\label{eq:eq1}
\end{equation}
where MSA\textsubscript{HRCHY\textsubscript{l}} denotes the multi-head self-attention layer of hierarchy $l$, $\hat{z}^{t}$ and $z^{t}$ are the output representations of MSA and MLP. In the practice, MSA\textsubscript{HRCHY\textsubscript{l}} is applied parallel to all partitioned blocks:

\begin{equation}
\begin{array}{l}
\text{MSA\textsubscript{HRCHY\textsubscript{l}}(Q, K, V)} =  \text{Stack}(\text{BLK}\textsubscript{1},..., \text{BLK}\textsubscript{T})\\ \text{BLK}= \text{Attention(Q, K, V)}=\text{Softmax}(\frac{QK^{\top}}{\sqrt{\sigma}})V,
\end{array}
\label{eq:eq2}
\end{equation}
where $Q, K, V$ denote queries, keys, and values vectors in the multi-head attention, $\sigma$ is the size of each vector. All blocks at each level of hierarchy share the same parameters given the input $\mathcal{X}$, which leads to hierarchical representations without increasing complexity. Finally, the block aggregation is merged spatially by adjacent 8 blocks.

\subsection{3D Block Aggregation}
Following~\cite{zhang2021nested}, we extend the spatial nesting operations to 3D blocks where each volume block is modeled independently. Information across blocks is communicated by the aggregation module. At hierarchy $l$, the spatial operations are conducted to down-sampled feature maps at $\mathbb{R}^{b \times H'/2 \times{W'/2} \times{D'/2}}$. At the bottom of each hierarchy, the embeddings are blocked back to feature $Z_{l+1} \in \mathbb{R}^{b \times{T/8} \times{n}}$ for hierarchy $l+1$. There are three hierarchies in our model design, a factor of 8 is reduced in a total number of blocks which results in [64, 8, 1] blocks. In the volumetric plane, the encoded blocks are merged among adjacent blocks representations. The design and use of the aggregation modules in the 3D scenario leverage local attention, lead to a data-efficient design. 

\subsection{Decoder}

To better capture localized information and further reduce the effect of lacking inductive bias of transformer, we use a hybrid design with a convolution-based decoder for segmentation. The features from different hierarchies of the transformer encoder are fed into skip connections followed by convolution layers. As shown in Figure~\ref{fig:fig2}, we extract the output representations at the image level and each hierarchy to $3 \times 3 \times 3$ conv layers, then upsample by a factor of $2$. Next, the output of the transposed conv is concatenated with the prior hierarchy representations. The segmentation mask is acquired by $1 \times 1 \times 1$ conv layer with a softmax activation function. Compared to some prior related works such as TransBTS~\cite{wang2021transbts} and CoTr~\cite{xie2021cotr}, our design employs the hierarchical transformer directly on images and extract representations at multiple scales without conv layers.

\section{Experiments}
\subsection{Dataset}
\label{dataset}
\noindent\textbf{Renal Substructure Dataset.}
The study design uses clinically collected renal CT of 116 de-identified patients accessed under IRB approval. We use selected ICD codes related to kidney dysfunction as exclusion criteria, that could have a potential influence on kidney anatomies. The left and right renal structures are outlined manually by three interpreters under the supervision of clinical experts. The annotation for the cortex label also includes the renal columns, the medulla is surrounded by the cortex, and the pelvicalyceal systems contain calyces and pelvis that drain into the ureter. All manual labels are verified and corrected independently by expert observers. For the test set of 20 subjects, we perform a second round of manual segmentation (interpreter 2) to assess the intra-rater variability and reproducibility.

\noindent\textbf{KiTS19.}
To validate the generalizability of the proposed method while remaining the target of characterizing renal tissues, we apply the model to the public KiTS19 dataset. The KiTS19~\cite{heller2021state} task focuses on the whole kidney and tumor segmentation. We perform five-fold cross-validation experiments and show results of the held-out 20\% as testing.

\subsection{Implementation Details}
\label{details}
Five-fold cross-validation is used for all experiments on 96 subjects, while 20 subjects are used for held-out testing. The five-fold models' ensemble is used for inferencing and evaluating test set performance. For experiment training, we used 1) CT window range of [-175, 275] HU; 2) scaled intensities of [0.0,1.0]; 3) training with single Nvidia RTX 2080 11GB GPU with Pytorch and MONAI implementation at batch size of 1 (input image sub-volume size of $96 \times{96} \times{96}$); 4) AdamW optimizer with warm-up cosine scheduler of 500 steps. The learning rate is initialized to 0.001 followed by a decay of $1e^{-5}$ for 50K iterations. For fair comparison and direct evaluation of the effectiveness of models, no pre-training is performed for all segmentation tasks.

\noindent\textbf{Metrics.}
Segmentation performance is evaluated between ground truth (rater 1) and prediction by Dice-Sorensen coefficient (DSC), and symmetric Hausdorff Distance (HD). Volumetric analyses are evaluated under R squared error, Pearson R, absolute deviation of volume, and the percentage difference between the proposed method and manual label.

\begin{table}[t]
\centering
\scriptsize
\setlength{\tabcolsep}{2mm}
\renewcommand\arraystretch{1}
\caption{Segmentation results of the renal substructure on testing cases. The UNesT achieves state-of-the-art performance compared to prior kidney components studies and 3D medical segmentation baselines. The number of parameters and GFLOPS (with a single input volume of $96 \times 96 \times 96$) are shown for deep learning-based approaches. * indicates statistically significant ($p < 0.01$) by Wilcoxon signed-rank test.}
\resizebox{1.0\textwidth}{!}{

\begin{tabular}{l|cc|cccccc|cc}
\hline
\multirow{2}{*}{Method}   &\multirow{2}{*}{\#Param} &\multirow{2}{*}{GFLOPS} &\multicolumn{2}{c}{Cortex}  &\multicolumn{2}{c}{Medulla} &\multicolumn{2}{c}{Pelvicalyceal System} &\multicolumn{2}{|c}{Avg.} \\
\cline{4-11}
 & & &DSC &HD &DSC &HD &DSC &HD &DSC &HD\\
\hline
Chen et al.~\cite{chen2012automatic} &N/A &N/A &0.7512 & 40.1947    &N/A &N/A    &N/A &N/A     &N/A &N/A    \\
Xiang et al.~\cite{xiang2017cortexpert} &N/A &N/A &0.8196 & 27.1455    &N/A &N/A    &N/A &N/A     &N/A &N/A    \\
Jin et al.~\cite{jin20163d} &N/A &N/A &0.8041 & 34.5170    &0.7186 &32.1059    &0.6473 &39.9125     &0.7233 &35.5118    \\
Tang et al.~\cite{tang2021renal} &40.9M &423.9 &0.8601 & 19.7508    &0.7884 &18.6030    &0.7490 &34.1723     &0.7991 &24.1754    \\
\hline
nnUNet~\cite{isensee2021nnu} &19.1M (3DUNet) &412.7 &0.8915 & 17.3764    &0.8002 &18.3132    &0.7309 &31.3501     &0.8075 &22.3466    \\
TransBTS~\cite{wang2021transbts} &33.0M &359.4	&0.8901 & 17.0213    &0.8013 &17.3084    &0.7305 &30.8745     &0.8073 &21.7347    \\
CoTr~\cite{xie2021cotr} &46.5M &399.2	&0.8958 & 16.4904    &0.8019 &16.5934    &0.7393 &30.1282     &0.8123 &21.0707    \\
nnFormer~\cite{zhou2021nnformer} &158.9M &146.5	&0.9094 & 15.5839    &0.8104 &15.9412    &0.7418 &29.4407     &0.8205 &20.3219    \\
UNETR~\cite{hatamizadeh2021unetr} &92.6M &41.2	&0.9072 & 15.9829    &0.8221 &14.9555    &0.7632 &27.4703     &0.8308 &19.4696    \\
\hline
UNesT  &87.3M &37.5	&\textbf{0.9201} &\textbf{14.5401}    &\textbf{0.8356} &\textbf{13.5933}    &\textbf{0.7843} &\textbf{24.5445}     &\textbf{0.8467*} &\textbf{17.5593}    \\
\hline
\end{tabular}
}
\label{tab:tab1}
\end{table}

\begin{table}[t]
\centering
\scriptsize
\setlength{\tabcolsep}{2mm}
\renewcommand\arraystretch{1}
\caption{Comparison of volumetric analysis metrics between the proposed method and the state-of-the-art clinical study on kidney components.}
\resizebox{1.0\textwidth}{!}{

\begin{tabular}{l|cc|cc|cc}
\hline
\multirow{2}{*}{Metrics}   &\multicolumn{2}{c}{Cortex}  &\multicolumn{2}{c}{Medulla} &\multicolumn{2}{c}{Pelvicalyceal System} \\
\cline{2-7}
 &Tang et al.~\cite{tang2021renal}  &UNesT &Tang et al.~\cite{tang2021renal} &UNesT &Tang et al.~\cite{tang2021renal} &UNesT 
\\
\hline
R Squared		&0.9200 & 0.9359    &0.6652 &0.6837    &0.4586 &0.5917       \\
Pearson R			&0.9838 & 0.9891    &0.8156 &0.8368    &0.6772 &0.7148     \\
Absolute Deviation of Volume &3.0233 & 2.7254    &3.5496 &3.2958    &0.9443 &0.8012      \\
Percentage Difference			&4.8280 & 3.9478    &7.4750 &7.0382    &19.0716 &13.5737   \\

\hline
\end{tabular}
}

\label{tab:tab2}
\end{table}
\section{Results}

\subsection{Characterization of Renal Structures}
We evaluate the UNesT performance on two groups of methods: 1) the clinical kidney components studies such as CortexSeg~\cite{chen2012automatic}, CorteXpert~\cite{xiang2017cortexpert}, AAM~\cite{jin20163d},  and 2) recent conv-~\cite{isensee2021nnu} and transformer-based~\cite{wang2021transbts,xie2021cotr,zhou2021nnformer,hatamizadeh2021unetr} 3D medical segmentation baselines. 

\begin{figure}[t]
\centering
\includegraphics[width=\textwidth]{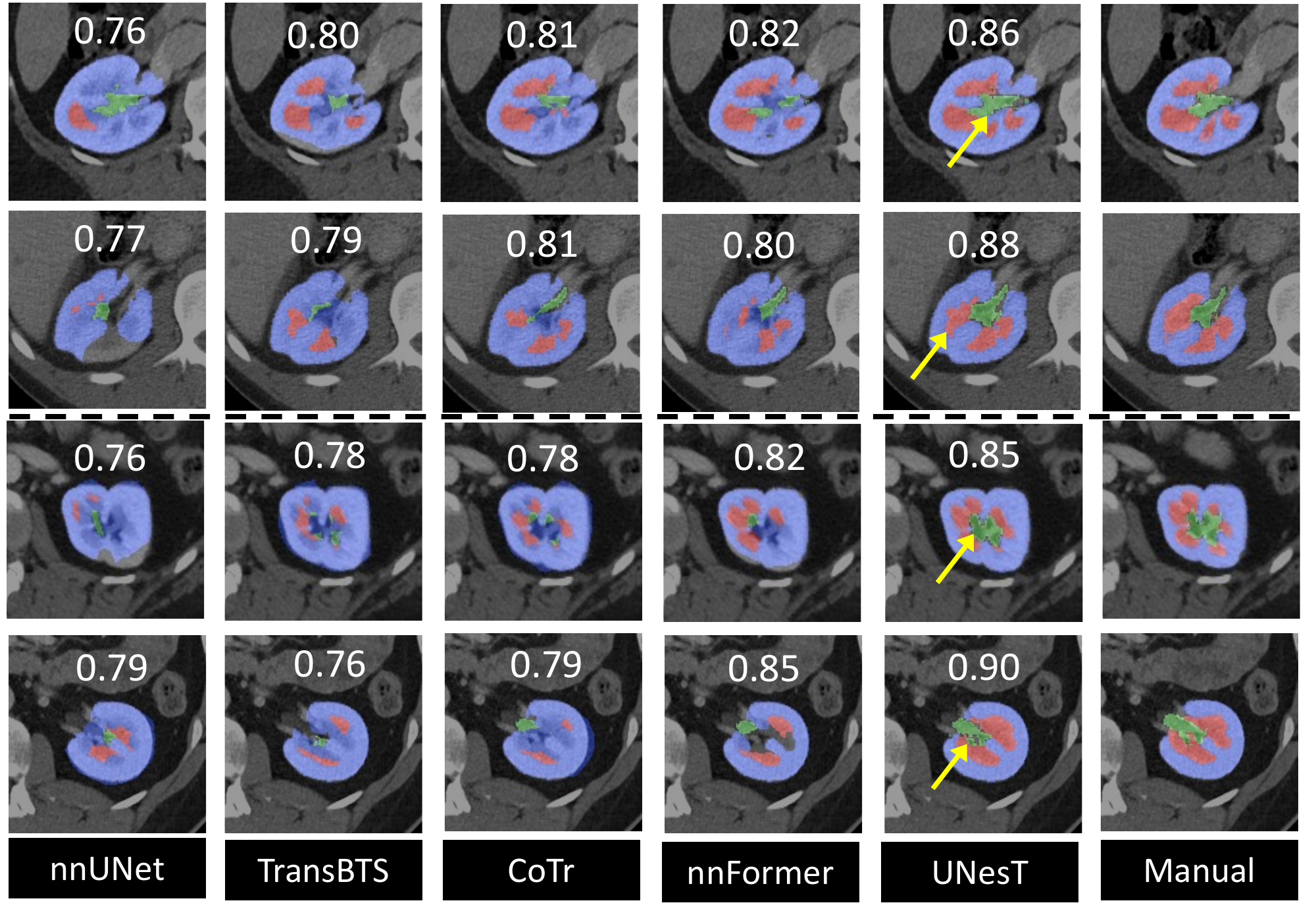}
\caption{Qualitative comparisons of representative renal sub-structures segmentation on two right (top) and two left (bottom) kidneys. The average DSC is marked on each image. UNesT shows distinct improvement on the medulla (red) and pelvicalyceal system (green) against baselines.}
\label{fig:fig3}
\end{figure}

\noindent {\bf Segmentation Results.}
Compared to canonical kidney studies using shape model or random forests in Table~\ref{tab:tab1}, the deep learning-based methods improve the performance by a large margin from 0.7233 to 0.7991. Among the nnUNet~\cite{isensee2021nnu} and extensive transformer models, we obtain the state-of-the-art average Dice score of 0.8467 compared to the second-best performance of 0.8308, with a significant improvement $p < 0.01$ under Wilcoxon signed-rank test. We observe higher improvement on smaller anatomies such as medulla and collecting systems. We compare qualitative results in Figure~\ref{fig:fig3}. Our method demonstrates the distinct improvement of detailed structures for medulla and pelvicalyceal systems.

\noindent {\bf Volumetric Analysis.}
Table~\ref{tab:tab2} lists the volume measurement with the proposed method. The UNesT achieves an R squared error of 0.9359 on the cortex. The correlation performance metric with Pearson R achieves 0.9891 for the UNesT against the manual label on the cortex. Our method obtains 2.7254 with an absolute deviation of volumes. The percent difference in the cortex is 3.9478. Quantitative results show that our workflow can serve as the state-of-the-art volumetric measurement compared to prior kidney characterization pipeline~\cite{tang2021renal}. 

\subsection{Ablation Study}
\begin{figure}[t]
\centering
\includegraphics[width=\textwidth]{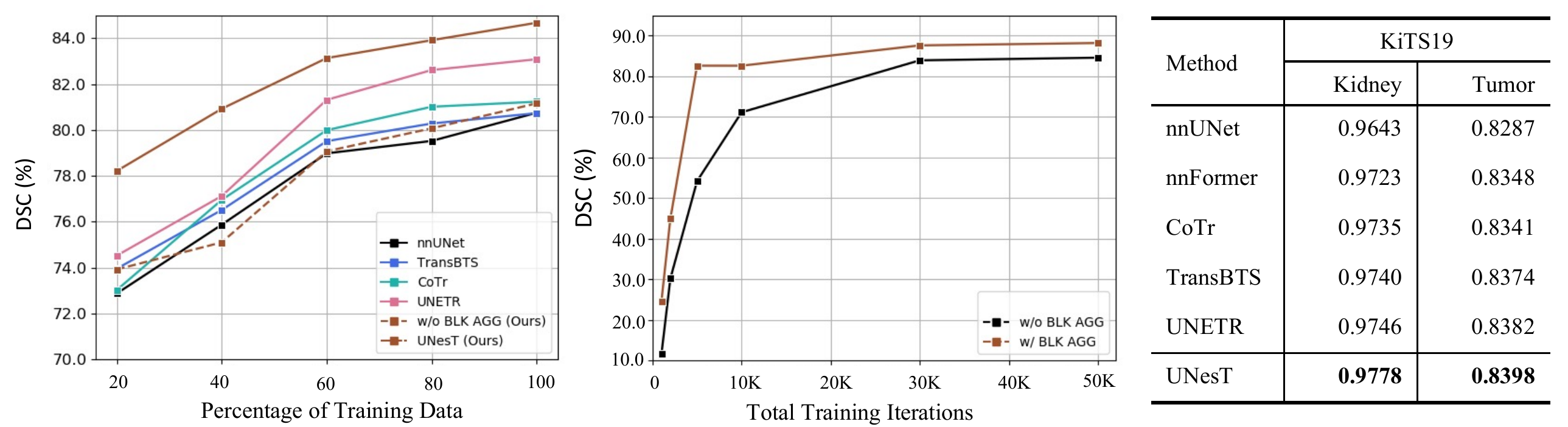}
\caption{Left: DSC comparison on the test set at different percentages of training samples. Middle: Comparison of the convergence rate for the proposed method with and without hierarchical modules, validation DSC along training iterations are demonstrated. Right: Results on the KiTS19 dataset show the generalizability of the proposed UNesT.}

\label{fig:fig4}
\end{figure}

\noindent {\bf Effect of the Block Aggregation.} We show the hierarchical architecture design (with 3D block aggregation) is critical for medical image segmentation (as shown in Figure~\ref{fig:fig4} left and middle). The result shows that the hierarchy mechanism achieves superior performance at 20\% to 100\% of training data. At the low data regime, the block aggregation achieves a higher improvement ($> 4\%$ of DSC) compared to the second-best method. We notice that the model without block aggregation (canonical transformer layers) obtains lower performance. The results show that block aggregation performs as a critical component for representation learning for transformer-based models.

\noindent {\bf Data Efficiency.} The Figure~\ref{fig:fig4} shows the data efficiency of our proposed method. First, UNesT achieves better performance when training with fewer data. Second, UNesT with block aggregation demonstrates a faster convergence rate (15\% and 4\% difference at 2K/30K iterations) compared to the backbone model without hierarchies.

\noindent {\bf Generalizability.} To validate the generalizability of the UNesT, we compare KiTS19 results among nnUNet~\cite{isensee2021nnu} and transformer-based methods. Our approach achieves moderate improvement at DSC of 0.9778 and 0.8398 for kidneys and tumors, indicating that the designed architecture can be used as a generic 3D segmentation method.

\section{Discussion and Conclusion}
In this paper, we target the critical problem that transformer-based models are commonly data-inefficient, which leads to unsatisfied performance when tasked with learning small structures and small datasets. In this work, we develop the first cohort of renal sub-structures study, specifically the renal cortex, medulla, and pelvicalyceal system. Upon the clinically acquired subjects, we propose a novel hierarchical transformer-based 3D medical image segmentation approach (UNesT). We show that the proposed method is data-efficient for accurately quantifying kidney components and can be used for volumetric analysis such as the medullary pyramids. Figure~\ref{fig:figA1} in the supplementary materials shows the proposed automatic segmentation method achieves better agreement compared to inter-rater assessment, 0.01 against 0.29 of mean difference indicating reliable reproducibility.

\begin{figure}[htp]
\centering
\includegraphics[width=\textwidth]{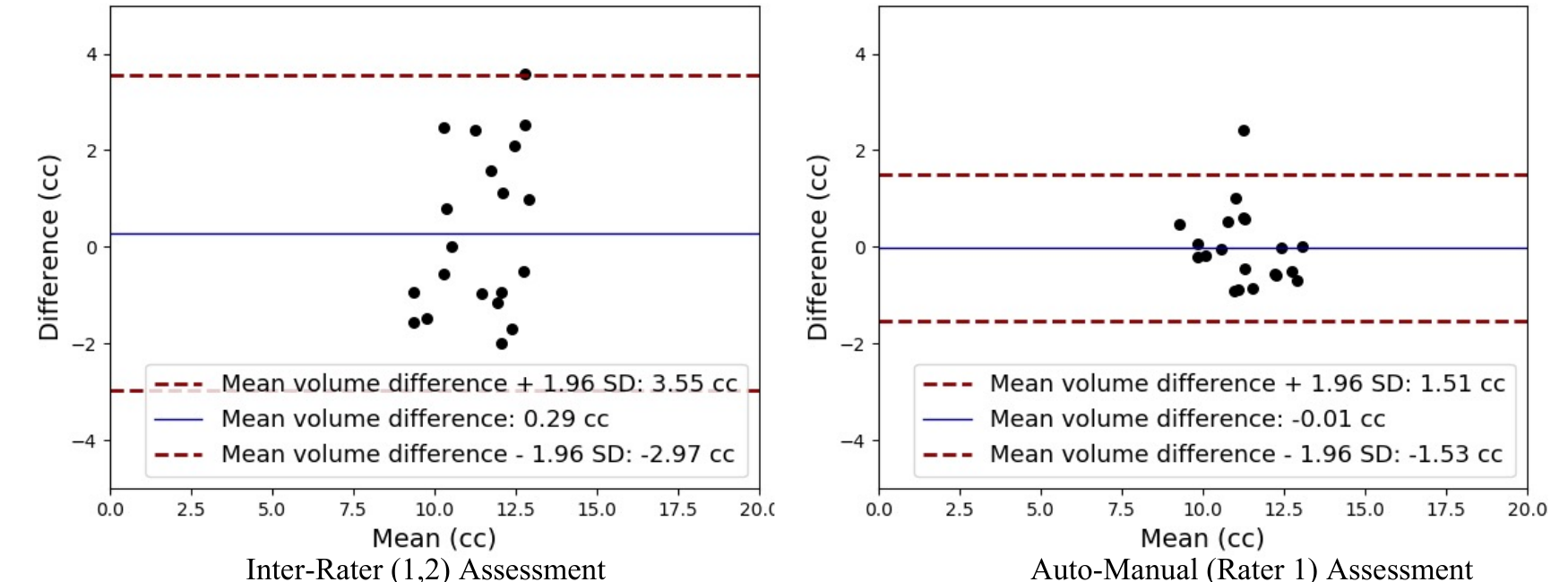}
\caption{The Bland-Atman plots compare the medulla volume agreement of inter-rater and auto-manual assessment. We show the cross-validation on interpreter 1, interpreter 2 manual segmentation on the same test set.  Interpreters present independent observation without communication. The auto-manual assessment shows the agreement between UNesT and interpreter 1 annotations. }
\label{fig:figA1}
\end{figure}

\noindent {\bf  Clinical Impact.}
Visual quantitative analysis of renal structures remains a complex task for radiologists. Some of the histomorphometry features of regions of the kidney (e.g. textural or graph features) are poorly adapted for manual identifications. In this study, we show that UNesT achieves consistently reliable performance. Compared with previous studies on cortex segmentation, the proposed approach significantly facilitates derivation of the visual and quantitative results.

Efficient segmentation is critical for clinical practice in deploying individual assessment. We note that, unlike other large organs, the renal segmentation dataset can be different in terms of imaging protocols, patient morphology, and institutional variations. We consider the framework adaptable to the segmentation of abnormal primitives in the future. In terms of sensitivity, we believe that the approach can be further improved from two perspectives. First, pre-registration of the kidney region of interest can help to reduce the shape and size variations and thus boost the segmentation performances. Second, incorporating dose usage in the segmentation loop can be very helpful. It can be expected that augmented contrast can be measured to better identify adjacent tissues among renal structures. \\

%
%
%
%

\noindent {\bf Acknowledgements.}
This research is supported by NIH Common Fund and National Institute of Diabetes, Digestive and Kidney Diseases U54DK120058, NSF CAREER 1452485, NIH grants, 2R01EB006136, 
1R01EB017230 (Landman), and R01NS09529. The identified datasets used for the analysis described were obtained from the Research Derivative (RD), database of clinical and related data. The imaging dataset(s) used for the analysis described were obtained from ImageVU, a research repository of medical imaging data and image-related metadata. ImageVU and RD are supported by the VICTR CTSA award (ULTR000445 from NCATS/NIH) and Vanderbilt University Medical Center institutional funding. ImageVU pilot work was also funded by PCORI (contract CDRN-1306-04869). We thank Ali, Vishwesh, Dong, Holger and Daguang at Nvidia for the 3D transformer discussions on 2021 summer. \\\\\\

\bibliographystyle{splncs04}
\bibliography{paper}

\end{document}